# Pressure-driven electronic and structural phase transition in intrinsic magnetic topological insulator $MnSb_2Te_4$

Yunyu Yin[1, †], Xiaoli Ma[1, †], Dayu Yan[1,2 †], Changjiang Yi[1], Binbin Yue[3], Jianhong Dai[1], Lin Zhao[1], Xiaohui Yu[1, 2, 4, *], Youguo Shi[1, 2], Jian-Tao Wang [1, 2, 4*], Fang Hong[1, 2, 4, *]

[1]*Beijing National Laboratory for Condensed Matter Physics and Institute of Physics, Chinese Academy of Sciences, Beijing 100190, China*

[2]*School of Physical Sciences, University of Chinese Academy of Sciences, Beijing 100190, China*

[3]*Center for High Pressure Science & Technology Advanced Research, Beijing, 100094, China*

[4]*Songshan Lake Materials Laboratory, Dongguan, Guangdong 523808, China*

[†]These authors have equal contribution.

* Email: yuxh@iphy.ac.cn; ygshi@iphy.ac.cn; wjt@iphy.ac.cn; hongfang@iphy.ac.cn

**Abstract**

Intrinsic magnetic topological insulators provide an ideal platform to achieve various exciting physical phenomena. However, this kind of materials and related research are still very rare. In this work, we reported the electronic and structural phase transitions in intrinsic magnetic topological insulator $MnSb_2Te_4$ driven by hydrostatic pressure. Electric transport results revealed that temperature dependent resistance showed a minimum value near short-range antiferromagnetic (AFM) ordering temperature $T_N'$, the $T_N'$ values decline with pressure, and the AFM ordering was strongly suppressed near 10 GPa and was not visible above 11.5 GPa. The intensity of three Raman vibration modes in $MnSb_2Te_4$ declined quickly starting from 7.5 GPa and these modes become undetectable above 9 GPa, suggesting possible insulator-metal transition, which is further confirmed by theoretical calculation. In situ x-ray diffraction (XRD) demonstrated that an extra diffraction peak appears near 9.1 GPa and $MnSb_2Te_4$ started to enter an amorphous-like state above 16.6 GPa, suggesting the structural origin of suppressed AFM ordering and metallization. This work has demonstrated the correlation among interlayer interaction, magnetic ordering, and electric behavior, which could be benefit for the understanding of the fundamental properties of this kind of materials and devices.

## Introduction

When magnetic ordering is introduced to topological insulators, they will display a lot of unusual physical properties, such as topological quantum behavior, Chern-insulator and axion-insulator phases, and name so-called “intrinsic magnetic topological insulators” [1-4]. However, this kind of materials are still very rare, $MnBi_2Te_4$ and $MnSb_2Te_4$ have been recently synthesized in forms of single crystal and epitaxial film, and studied by few groups [5-11]. The realization of the quantum anomalous Hall effect (QAHE) in Cr-doped (Bi, Sb)$_2$Te$_3$ successfully introduces magnetism into time-reversal-invariant topological insulators [12-17]. And the QAHE has also been realized by Deng *et al.* in odd-layered $MnBi_2Te_4$ at 1.4 K without magnetic field and it can be enhanced to 6.5 K by external magnetic field [2], which helps to align all magnetic layers in ferromagnetic geometry. Gong *et al.* investigated MBE-grown $MnBi_2Te_4$ films by SQUID and ARPES, observed the anisotropic magnetic behavior and the highly layer-number dependent behavior of electronic band structures [5]. The 7 septuple-layer shows archetypal Dirac-type energy bands with isotopic dispersion behavior and typical 3D character, quite different from the warped surface state of pure $Bi_2Te_3$ [5]. Shi *et al.* reported the magnetoresistance effect and anomalous hall effect (AHE) in mechanical exfoliated $MnSb_2Te_4$, and AHE can be observed until ~35 K [7].

$MnBi_2Te_4$ and $MnSb_2Te_4$ are sister materials and form in same structure (Fig. 1), while $MnBi_2Te_4$ is electron-carrier semiconductor and $MnSb_2Te_4$ is a hole-carried semiconductor [18-22]. Both can be mechanically exfoliated by special procedure and the interlayer interaction is quite similar with the van-der-Waals interaction in $Bi_2Te_3$, which are sensitive to external strain effect [23, 24]. External pressure could help to reveal the relation between interlayer interaction and fundamental properties [25-28], which is benefit for effective controlling of their exotic properties and promoting the application in field of multifunctional electronic devices. Pei *et al.* have done systematic study on $MnBi_2Te_4$ by using diamond anvil cell (DAC) [24], and found that it undergoes a metal-semiconductor-metal transition under high pressure and its AFM ordering is suppressed. A similar work was done by Chen *et al.* with lower pressure range and displayed similar results [23]. The competition between bulk state and surface state is supposed to be responsible for the two step transitions on resistivity [24].

Since the carrier type in $MnSb_2Te_4$ is different from $MnBi_2Te_4$, how the electronic properties response to external pressure/strain is of great scientific interesting [29-31]. In this work, we did systematic study on $MnSb_2Te_4$ under high pressure, based on electric transport, in situ Raman spectroscopy, synchrotron XRD and theoretical calculation. $MnSb_2Te_4$ undergoes a phase transition from original crystalline phase to an amorphous-like phase upon compression, and experience the semiconductor-poor metal-good metal transition, which is quite different from the phase change route observed in $MnBi_2Te_4$. There is a strong anisotropic behavior of out-of-plane Sb/Mn-Te interaction, revealed by the anisotropic compression behavior of two phonon modes.

Meanwhile, the short-range ferrimagnetic ordering is suppressed with pressure and is finally destroyed near 17 GPa, which is well consistent with the amorphous process. This work has demonstrated that the electronic behavior is totally different in these two systems with hole and electron carrier, respectively. Short-range spin ordering in $MnSb_2Te_4$ could arouse strong scattering behavior on electric transport and it always exists in crystalline phase, which provides a good platform to study the correlation behavior of spin and charge in intrinsic magnetic topological insulators.

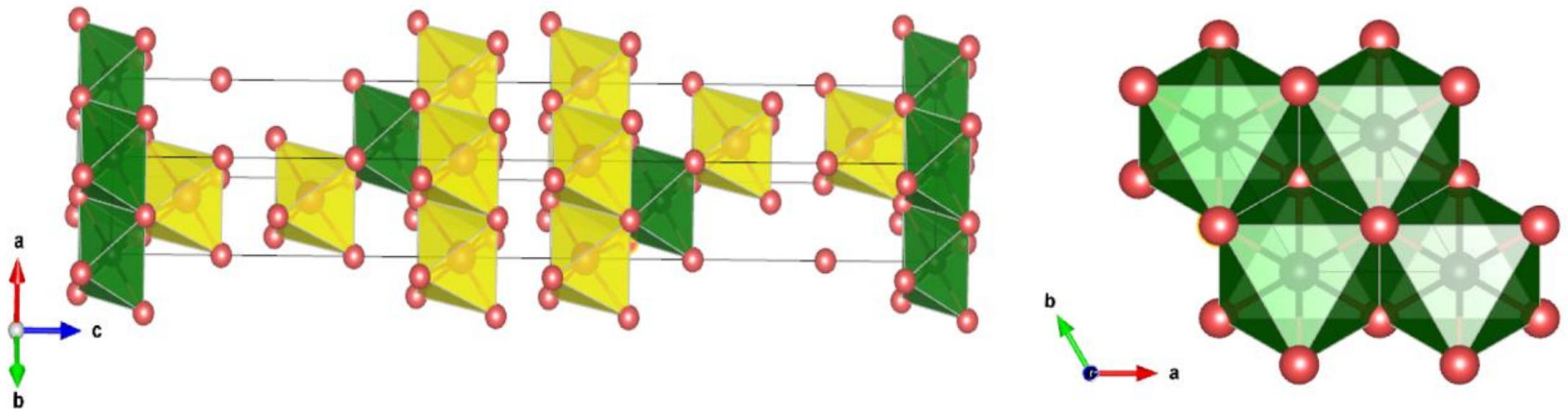

**Fig. 1** The atomic structure of intrinsic magnetic topological insulator $MnSb_2Te_4$. The yellow octahedral is $SbTe_6$ unit while green octahedral is $MnTe_6$ unit.

**Experiment**

The $MnSb_2Te_4$ single crystals were synthesized via the flux method. High purity raw materials Mn (99.99%), Sb (99.9999%) and Te (99.9999%) were ground and mixed at a molar ratio of 1:10:16 in glove box with inert gas protection. The mixture was then sealed in an evacuated quartz tube and heated to 700◦C for over 10 h and dwelt for 20 h in a furnace. A 300-micron culet DAC (BeCu) was used to produce hydrostatic environment. Re gasket was used as a support, c-BN epoxy mixtures were used as insulating materials and KBr was used as the pressure medium. Pressure is determined by the R1-R2 line shift of a ruby ball [32]. A thin and rectangle flake sample was placed in the center of the chamber and the electric transport measurements were carried out in a standard four-probe geometry inside a Janis cryostat, which is cooled by a closed-loop $H^4$ cycling system. In situ XRD data was collected with a wavelength of 0.6199Å at Beamline 4W2 at Beijing Synchrotron Radiation Facility.

Our density functional theory (DFT) calculations are performed using the Vienna *ab initio* simulation package (VASP) [33] with the projector augmented wave method [34] and spin polarized generalized gradient approximation [35] for the exchange-correlation energy. The valence states $3p^63d^64s^1$ for Mn, $5s^25p^3$ for Sb, and $5s^25p^4$ for Te are used with the energy cutoff of 500 eV for the plane wave basis set. To simulate the interlayer antiferromagnetic (AFM) coupling [36], an AFM rhombohedral primitive cell [37] is used with a van der Waals correction [38]. The Brillouin zone is sampled with a 7 × 7 × 7 Monkhorst-Pack special k-point grid. Throughout our calculations up to 15 GPa, the DFT + U method [39] is used with an effective value $U_{eff}$ = 3 eV for the Mn *3d* electrons. The geometries are optimized with symmetry constraints until the remaining atomic forces are less than $10^{-2}$ eV/Å and the energy convergence criterion is set at $10^{-8}$ eV. The electronic band structures and density of states are calculated by the modified Becke-Johnson (mBJ) functional [40] with the DFT + U method. The magnetic moments on Mn sites are calculated to be 4.42-4.45 $\mu_B$, and the 3*d* bands of Mn atoms are mainly located in an energy range between -5 and -6 eV below the Fermi level.

**Results and discussion**

The electric transport measurement was carried out based on standard four-probe geometry (Fig. 2(a)). Compared with the van der Pauw four-probe configuration, standard four-probe configuration provides a higher liability and stability on transport data collecting. At low pressure, ~1.6 GPa, the crystal shows metallic behavior above ~52 K and then turn into a semiconducting state (as seen in Fig. 2(a)), which is well consistent with the behavior of single crystal $MnSb_2Te_4$ at ambient condition [36]. This transition is assumed to associate with the short-range AFM correlation /ordering of Mn sub-lattice [36], while it shows a lower long-ordering transition temperature near 19 K [36], 25K [41] or 31 K [42] confirmed by the magnetic measurement. The difference on the transition temperature of long-range spin ordering in $MnSb_2Te_4$ indicates that the transition is sensitive to the site mixing, which cannot be well controlled during sample synthesis due to the close atomic size of Mn and Sb. As pressure increases, the temperature dependent resistance shows monotonous decreasing trend until 16.5 GPa (Fig. 2(a)). Meanwhile, the AFM short-range ordering temperature shifts towards lower temperature, which was strongly suppressed near 10 GPa and becomes undetectable above 11.5 GPa, above which the sample shows complete metallic behavior over the entire temperature range. However, the sample resistance shows abnormal behavior between 16.5 GPa and 21.0 GPa, during which the resistance at low temperature increases (Fig. 2(b)). Starting from 21.0 GPa, the temperature dependent resistance shows monotonous decreasing trend and doesn’t change too much above 31.0 GPa (Fig. 2(c)). For reference, the pressure dependent resistances (log scale) at 300 K and 5 K are plotted in the insets in Fig. 2 (b) and 2 (c), respectively.

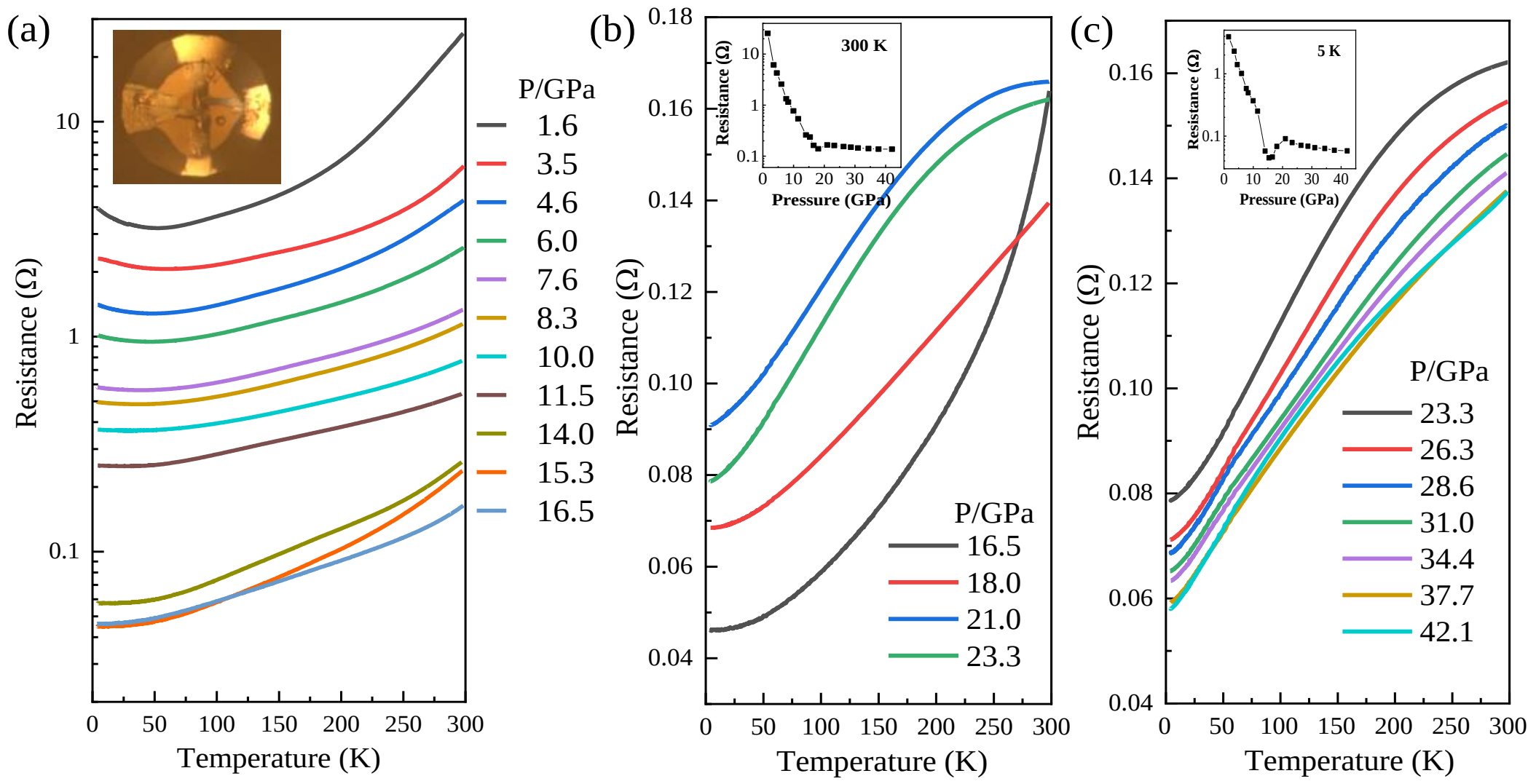

**Fig. 2** The electric transport properties (R-T curves) of $MnSb_2Te_4$ crystal under high pressure. (a) pressure up to 16.5 GPa, (b) up to 23.3 GPa, (c) up to 42.1 GPa, insets: (a) photograph of sample inside DAC chamber at 42.1 GPa, (b) the pressure dependent resistance at 300 K, (c) the pressure dependent resistance at 5 K.

Compared with high pressure transport behavior of $MnBi_2Te_4$ crystal [24], $MnSb_2Te_4$ shows some similarity but there is also much different behavior. Pei *et al.* and Chen *et al.* have done high pressure study on $MnBi_2Te_4$ independently [23, 24], their high-pressure experiments both demonstrated the suppression of AFM behavior, which is to some extent similar with our current work. Previous study demonstrates that $MnBi_2Te_4$ shows typical AFM ordering [2, 5, 18] while $MnSb_2Te_4$ shows ferrimagnetic ordering [42]. The ferrimagnetic behavior of $MnSb_2Te_4$ is mainly due to the anti-site behavior of Mn and Sb, which has been confirmed by neutron scattering experiment [43]. However, the transport behavior is totally different below this critical pressure. The resistance of $MnBi_2Te_4$ increases with pressure up to 10.3 GPa or 12.5 GPa [23, 24], while the resistance of $MnSb_2Te_4$ always decreases with pressure below 16.5 GPa. In addition, the original metallic state of $MnBi_2Te_4$ at low pressure become semiconducting state as pressure increase, and then turns back to metallic state above 19 GPa [24].

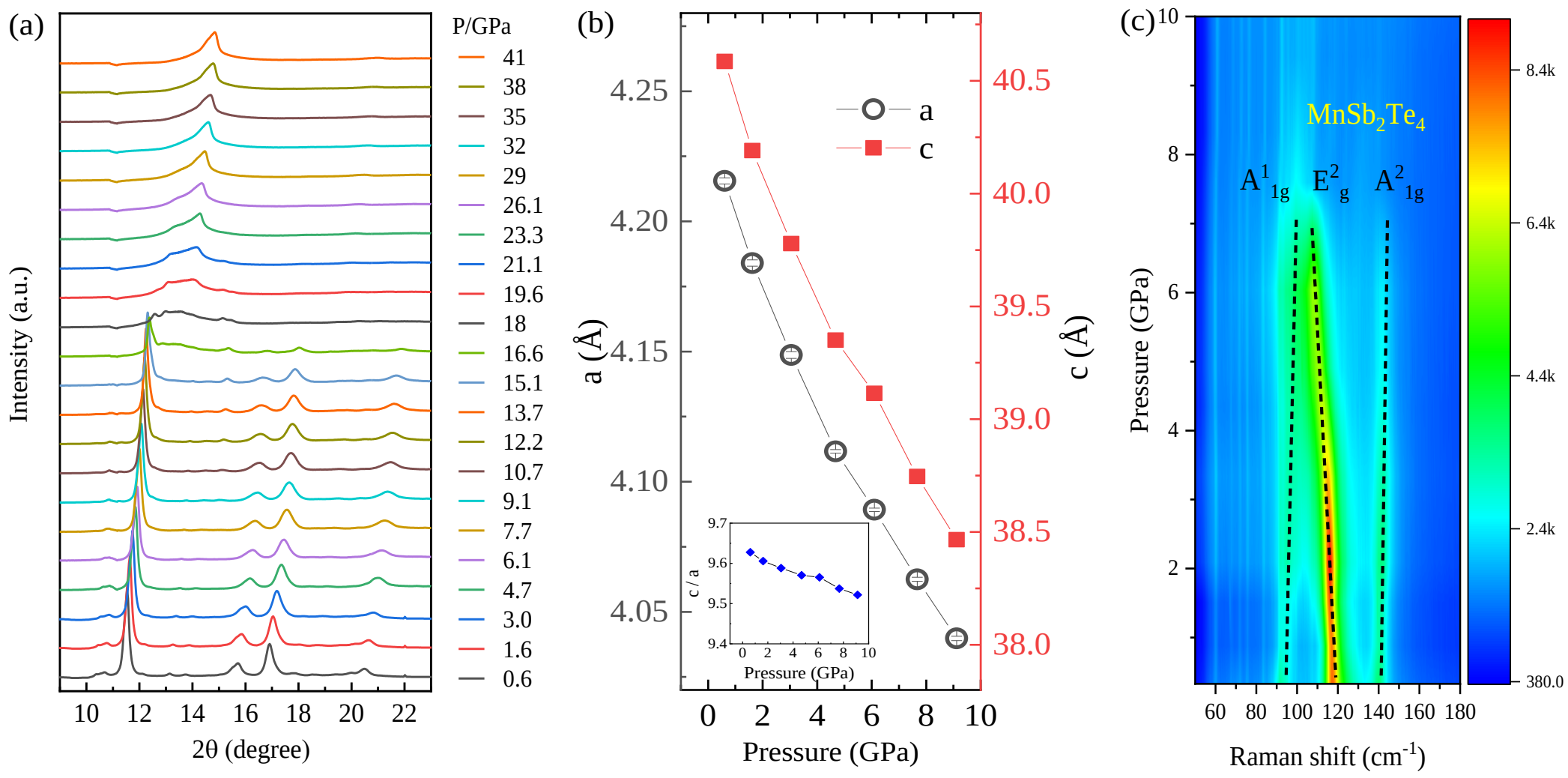

**Fig. 3** High pressure structural behavior. (a) in situ x-ray diffraction (XRD): a small peak appears near 2θ = 15°, starting from 9.1 GPa; Near 18 GPa, the sample starts to transform to an amorphous-like state; (b) the lattice parametrs up to 9.1 GPa, inset: *c/a* ratio, the change is very weak, *c/a* value at 9.1 GPa is only ~1% smaller than that at 0.6 GPa; (c) micro-Raman spectroscopy: three vibration modes were observed; there is strong signal drop near 7.5 GPa and it is undetectable above 9 GPa; the mode near 120 $cm^{-1}$ shows negative compression behavior.

In order to understand the underlying mechanism of the suppressed magnetism in $MnSb_2Te_4$ and the electric transport anomaly upon compression, we did in situ study on the static and dynamic structures by using synchrotron XRD and Raman spectroscopy, as presented in Fig. 3. Based on the XRD data under various pressures, we can observe a clear phase transition between 16.6 and 18.0 GPa, during which the intensity of most diffraction starts to become weaker and weaker (Fig. 3(a)). The sample itself finally enter an amorphous-like phase upon further compression, as shown in Fig. S1 and S2 in Supporting Information. Another small change is that there is new peak appearing starting at 9.1 GPa near 2θ ≈ 15°. All the other diffraction peaks of original phase are kept. To reveal the possible phase transition near 9.1 GPa, we did theoretical simulation to search possible new phases under high pressure. However, even when we relaxed the crystal lattice parameter and the stacking form of different layers, the final stable structure always went back to the original phase. Pei *et al.* did systematic study on $MnBi_2Te_4$ under high pressure and they only observed the amorphous transition and didn't find any extra anomaly at low pressure range [24]. Hence, we speculate that the new peak could be contributed by a superstructure or local short-range ordering. Similar transition was also observed in $SnBi_2Te_4$ [44]. Unfortunately, the accurate structure cannot be solved due to the low energy x-ray or short *q* range in current work and in $SnBi_2Te_4$ as well [44], and further work is still required to solve the fine structure above 9.1 GPa. The lattice parameters for original *R-3m* phase obtained from XRD pattern refinement is presented in Fig. 3(b). The pressure dependent *c/a* ratio shown in the inset of Fig. 3(b), and the change is quite small. At 9.1 GPa, the *c/a* ratio is only reduced by ~1% compared with the value at 0.6 GPa, suggesting the isotropic compressibility of $MnSb_2Te_4$, though it is in form of a layered structure. The raw data of lattice parameters was provided in Table S1 in Supporting Information.

To reveal the lattice dynamic behavior, Raman spectra were collected in situ inside (DAC). Three vibration modes have been observed and the initial position of three modes are at ~92, 118, and 140 $cm^{-1}$ (Fig. 3(c)). Upon compression, two of them shifts to higher wavenumber while the middle mode shows negative compression behavior. The intensity of all Raman modes becomes weak above 7.5 GPa and undetectable above 9.0 GPa. The mode at ~92 $cm^{-1}$ is still a mysterious mode, Wang *et al.* had been assigned to a surface phonon mode (~95 $cm^{-1}$) [45], and Guo *et al.* claimed the effect of surface oxidation [46], Rodríguez-Fernández *et al.* found a peak of ~88 $cm^{-1}$ in Te cluster in $Bi_2Te_3$ with rich Te content (which is similar with the ~95 $cm^{-1}$) [47], and Mal *et al.* found a ~94 $cm^{-1}$ mode in their calculation but didn't provide detailed information about the origin of this mode [48]. In the work of Mal *et al.*, the ~94 $cm^{-1}$ mode kept at the same position with temperature, while other modes show clear temperature dependence [48]. In Pei *et al.*' work on $MnBi_2Te_4$, they didn't observe any vibration mode near between 80 and 100 $cm^{-1}$ [24]. Hence, more theoretical work is required to explain the origin of the ~92 $cm^{-1}$ in $MnSb_2Te_4$. The 118 $cm^{-1}$ mode is expected to be an IR-active mode ($A^1_{1u}$) in bulk [48], but it has been activated by symmetry break (may be due to surface effect of laser penetration or anti-site effect of Mn and Sb) and observed by Raman spectroscopy as a $E^2_g$, corresponding to the out-of-plane vibration of Sb (and/or anti-site Mn) and Te atoms. The 118 $cm^{-1}$ displays a pressure negative dependence behavior, which is rarely reported in layered materials. The 142 $cm^{-1}$ mode can be assigned to $A^2_{1g}$ mode [48], which corresponds to out-of-plane vibration of Sb (and/or anti-site Mn) and Te atoms as well. The different pressure dependence behavior of 118 and 142 $cm^{-1}$ modes suggests that there is strong anisotropy interaction among out-of-plane Sb (and/or anti-site Mn) and Te atoms.

Based on the structural information revealed by XRD, the changes of resistance are well consistent with the structural evolution. The anomaly in resistance measurement between 16.5 and 18.0 GPa should origin from the structural phase transition starting at 16.6 GPa. Further compression induces amorphous-like behavior while the resistance doesn't change too much anymore, especially for those above 25.5 GPa. The electronic band structures and density of states (DOS) are calculated by the modified Becke-Johnson (mBJ) functional [40]. The electronic band structures and the partial DOSs for Mn, Sb and Te atoms are plotted in Fig. 4 at ambient condition, 5 GPa, and 10 GPa. The total DOS under high pressure was presented in Fig. S3. The magnetic moments on Mn sites are calculated to be 4.42-4.45 $\mu_B$, and the 3*d* bands of Mn atoms are mainly located in an energy range between -5 and -6 eV below the Fermi level. At ambient condition (~ 0 GPa), $MnSb_2Te_4$ is a typical semiconductor with a direct band gap of 0.298 eV. The valence band maximum (VBM) and conduction band minimum (CBM) are located at the Γ point, as seen in Fig. 4(a), and the VBM is mainly contributed by Te-*p* electrons, while the CBM is contributed by Sb-*p* and Te-*sp* electrons [see Fig. S3(a)]. With increasing of pressure up to 5 GPa, it becomes to an indirect band gap semiconductor with a gap of 0.098 eV [see Fig. 4(b)]. The VBM is located at the Γ point and mainly contributed by Te-*p* electrons, while the CBM is shifted along the Z-F direction and mainly contributed by Sb-*p* electrons and partial Te-*p* and Mn-*d* electrons [see Fig. S3(b)]. At 7 GPa, it has a small indirect band gap of 0.002 eV [see Fig. S3(c)]. Above 7 GPa, it becomes metallic, and the valence bands around the Γ point are crossing at the Fermi level. Under high pressure at 10 GPa, more valence and conduction bands are connected around the Fermi level, as seen in Fig. 4(c), showing a strong metallic behavior with a peak of DOS at Fermi level [see Fig. S3(d)], and the states around the Fermi level are mainly contributed by the Sb-*p* (56%) and partial Te-*p* (24%) and Mn-*d* (20%) electrons. The calculation results are consistent with the Raman data, though the critical pressure points for metal-insulator transition is a little bit different by ~ 1 GPa. As mentioned above, the Raman signal becomes extremely weak and even undetectable above 9.0 GPa, which is therefore due to the metallization of $MnSb_2Te_4$. Above 7.5 GPa, the Raman signal drops significantly and it should correspond to the appearance of “bad” metal state (as the bandgap is just close), compared with the “good” metal state above 9.0 GPa. Recently, we noted that H. Zhang *et al.* reported the pressure effect on $MnSb_2Te_4$ by theoretical calculation (the lattice parameters used in their work is a little bit different from ours), and the band gap (without spin-orbit coupling effect) of $MnSb_2Te_4$ was also suppressed by pressure [30], which is to some extent consistent with our calculation. A phase diagram was proposed based on above measurement and discussion for reference, as seen in Fig. S4 in Supporting Information.

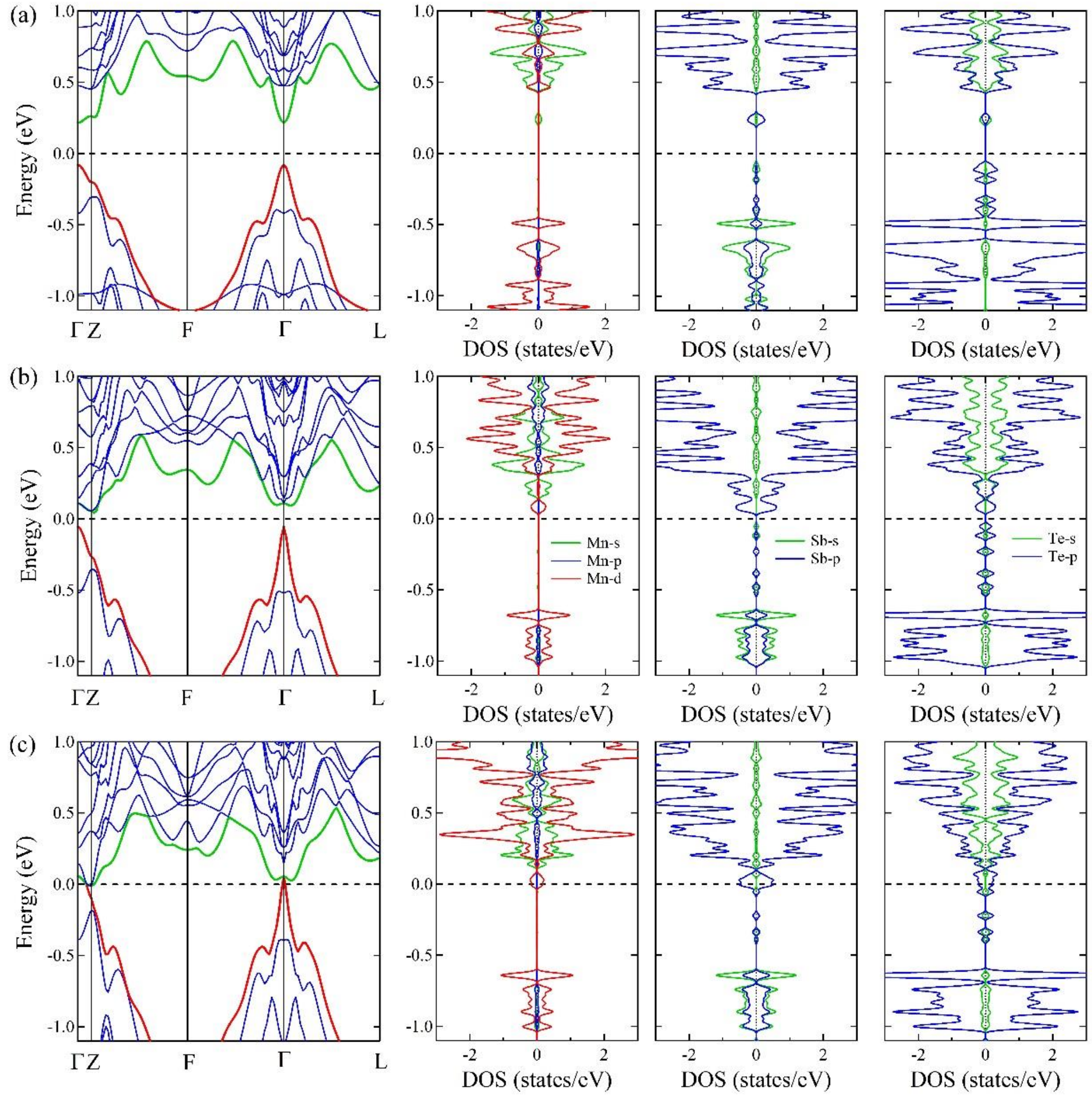

**Fig. 4** Calculated electronic band structures and partial DOS of up spins and down spins for $MnSb_2Te_4$ at (a) ambient condition (~ 0 GPa), (b) 5 GPa, and (c) 10 GPa using an AFM rhombohedral primitive cell. The lowest conduction band and highest valence band in band structures are marked with green and red lines. The direct band gap at ambient condition is 0.298 eV at Γ point. The *3d* bands of Mn atoms are mainly located in an energy range between -5 and -6 eV below the Fermi level and cannot see in this figure.

## Conclusion

In summary, we report the electric transport behavior and structural phase transition of intrinsic magnetic topological insulator $MnSb_2Te_4$ under high pressure up to ~ 42 GPa. At ambient pressure, electric resistance measurement demonstrated that $MnSb_2Te_4$ experiences a metal-semiconductor transition near 53 K, below which a short-range ferrimagnetic ordering is expected to form and is responsible for the resistance anomaly. Strong suppression of Raman signal above 7.5 GPa indicated the initiation of metallization and bandgap close, which is supported by the theoretical calculation. Clear metallization behavior is expected to exist at ~10 GPa by calculation, which is also consistent with the appearance of new structure and disappearance of any Raman signal above 9.1 GPa. It is also consistent with the full metallization above 11.5 GPa. The short-range ferrimagnetic ordering was sensitive to pressure and fully suppressed above 11.5 GPa, and $MnSb_2Te_4$ underwent a structural phase transition starting at 16.5 GPa and entered amorphous-like state. This work shows that the hole-carrier dominated $MnSb_2Te_4$ has a totally different electric behavior from the electron-carrier dominated $MnBi_2Te_4$ while their structure evolution under high pressure was very similar. Current work provides an insight to show the correlation among interlayer interaction, magnetic ordering, and electronic behavior.

**Author contribution**

F. Hong and X.H. Yu conceived the project. D. Y. Yan, C. J. Yi, Y. G. Shi synthesized the crystal. Y.Y. Yin and X.L. Ma did the resistance and Raman measurement. B. B. Yue and J. H. Dai carried out synchrotron x-ray diffraction experiment. J. T. Wang did the theoretical calculation. F. Hong loaded the samples for resistance, Raman, and synchrotron x-ray diffraction experiments, and wrote the manuscript. All authors joined the discussion and made comment on the manuscript.

**Acknowledgment**

This work was supported by the National Key Research and Development Program of China (Grant No. 2016YFA0401503, 2018YFA0305700 and 2020YFA0711502), the National Natural Science Foundation of China (Grant No. 11575288, 11974387, 12004416, 12004014 and 22090041), the Strategic Priority Research Program and Key Research Program of Frontier Sciences of the Chinese Academy of Sciences (Grant Nos. XDB33000000, XDB25000000 and Grant No. QYZDBSSW-SLH013) and the Youth Innovation Promotion Association of Chinese Academy of Sciences under Grant No. Y202003. Synchrotron x-ray diffraction experiment was done at Beamline 4W2 at Beijing Synchrotron Radiation Facility. Some instruments used in this study were built for the Synergic Extreme Condition User Facility.